\documentclass[pra,superscriptaddress,twocolumn,amsmath,amssymb]{revtex4-1}	
\usepackage{graphicx}
\usepackage{dcolumn} 
\usepackage{bm}
\usepackage{epsfig}
\usepackage{braket}
\usepackage{hyperref}
\hypersetup{
    colorlinks=true,        
    linkcolor=blue,          
    citecolor=blue,        
    filecolor=magenta,      
    urlcolor=blue           
}

\newcommand{\citenamefont}{}
\newcommand{\bibnamefont}{}

\usepackage{color} 				
\usepackage[normalem]{ulem} 			

\begin{document}
\title{Breaking of SU(4) symmetry and interplay between strongly correlated phases in the Hubbard model}

\author{A. Golubeva}
\affiliation{Institut f\"ur Theoretische Physik, Goethe-Universit\"at, 60438 Frankfurt/Main, Germany}
\affiliation{Perimeter Institute for Theoretical Physics, Waterloo, Ontario, N2L 2Y5, Canada}
\author{A. Sotnikov} \email{sotnikov@ifp.tuwien.ac.at}
\affiliation{Akhiezer Institute for Theoretical Physics, NSC KIPT, 61108 Kharkiv, Ukraine}
\affiliation{Institute of Physics, ASCR, Na Slovance 2, 182 21 Praha 8, Czech Republic}
\affiliation{Institute of Solid State Physics, TU Wien, Wiedner Hauptstr. 8, 1040 Wien, Austria}
\author{A. Cichy}
\affiliation{Institut f\"ur Theoretische Physik, Goethe-Universit\"at, 60438 Frankfurt/Main, Germany}
\affiliation{Institut f\"{u}r Physik, Johannes Gutenberg-Universit\"{a}t Mainz, Staudingerweg 9, D-55099 Mainz, Germany}
\author{J. Kune\v{s}}
\affiliation{Institute of Physics, ASCR, Na Slovance 2, 182 21 Praha 8, Czech Republic}
\affiliation{Institute of Solid State Physics, TU Wien, Wiedner Hauptstr. 8, 1040 Wien, Austria}
\author{W. Hofstetter}
\affiliation{Institut f\"ur Theoretische Physik, Goethe-Universit\"at, 60438 Frankfurt/Main, Germany}

\date{\today}

\begin{abstract}
  We study the thermodynamic properties of four-component fermionic mixtures described by the Hubbard model using the dynamical mean-field-theory approach. Special attention is given to the system with SU(4)-symmetric interactions at half filling, where we analyze equilibrium many-body phases and their coexistence regions at nonzero temperature for the case of simple cubic lattice geometry. We also determine the evolution of observables in low-temperature phases while lowering the symmetry of the Hamiltonian towards the two-band Hubbard model. This is achieved by varying interflavor interactions or by introducing the spin-flip term (Hund's coupling). By calculating the entropy for different symmetries of the model, we determine the optimal regimes for approaching the studied phases in experiments with ultracold alkali and alkaline-earth-like atoms in optical lattices.
\end{abstract}
\maketitle

\section{Introduction}
Mixtures of quantum particles with high spin symmetry in lattice systems attract significant attention in the scientific community for many reasons.
In particular, experimental realizations of systems that are invariant under continuous SU($N>2$) transformations can give valuable insight into mechanisms of spontaneous symmetry breaking that play a crucial role in vast areas of condensed-matter and high-energy physics. 
Depending on the specific symmetry, these systems are predicted to have rich phase diagrams and unique physical characteristics that are not yet fully explored.
A key property of multiflavor mixtures described by the Hubbard model is their high entropy capacity that constitutes a substantial advantage for approaching low-temperature many-body quantum phases experimentally with ultracold atoms in optical lattices.

Even though typical experimentally accessed temperatures and entropies of trapped atomic gases are too high 
to explore the realm of exotic low-temperature phases, experiments with multiflavor fermionic mixtures of $^6$Li, $^{40}$K, $^{87}$Sr, and $^{173}$Yb in optical lattices have already uncovered very rich physics of these systems~\cite{Caz2014RPP,Ott2008PRL,Tai2010PRL,DeSalvo2010PRL,Kra2012Nat,Tai2012Nat,Cappellini2012PRL, Pagano2014NP,Kra2014S,Scazza2014NP,Hof2016PRX}.
Of particular interest among these phases are N\'eel-type magnetically ordered phases and the metal-to-Mott-insulator transition.
Note that in the case of high spin symmetry (in contrast to large-$S$ representations of SU(2) symmetry in solid-state materials), quantum fluctuations increase with the number of components~\cite{Wu2012NP,Wu2006MPL}.

According to theoretical studies~\cite{Caz2014RPP,Gorshkov2010NP}, the N\'eel-type ordering is the dominant instability at half filling in SU($N$)-symmetric Hubbard models with up to $N=4$ interacting flavors. 
Mixtures with $N\gtrsim6$ start to favor nonmagnetic valence-bond (or, more generally, valence-cluster) states that govern the low-temperature physics of these systems.
Therefore, one can draw two main conclusions at this stage: With the increasing number of flavors, in general, (i) the entropy capacity increases and (ii) magnetic ordering is suppressed.
The notion of optimal parameter regime can be helpful to approach magnetically ordered states in experiments with a tunable number of fermionic flavors. From this perspective, four-component mixtures can be viewed as a promising candidate. Therefore, we choose this system for a detailed theoretical analysis of thermodynamic properties and relevant physical observables in the proximity of phase transitions.
A number of theoretical approaches have been developed recently to understand the low-temperature physics in the SU(4)-symmetric and other relevant four-component Hubbard and Heisenberg models. 
In particular, significant progress has been made in recent studies by
  quantum Monte Carlo (QMC) methods~\cite{Wu2006MPL,Cor2011PRL,Cai2013PRL,Cai2013PRB,Zhou2014PRB},
  dynamical mean-field theory (DMFT)~\cite{Inaba2005PRB,Blumer2013PRB,Hoshino2016PRB,Cichy2016PRA,Yanatori2016PRB},
  one-dimensional approaches~\cite{Capponi2016AP,Decamp2016PRA},
  high-temperature series expansion~\cite{Haz2012PRA}, and other mean-field approaches~\cite{Hof2004PRL,Che2007PRL,Szi2011EPL}.

\section{Model and method}
We describe the four-component interacting fermionic mixture in the framework of the well-known two-orbital Hubbard model with two internal spin states.
Besides the usual on-site intraorbital interaction $U$, there are two types of interorbital interactions denoted as {\it direct} ($V_{dir}$) and {\it exchange} ($V_{ex}$) interaction, respectively. 
Combining the spin and orbital indices into a single {\it flavor} index $\alpha$, the system is described by the general Hamiltonian
\begin{align} \label{eq:HubbardHwVex}
{\cal H} = &-t \sum \limits_{\langle i,j \rangle} \sum \limits_{\alpha = 1}^{4} (c_{i \alpha}^{\dagger} c_{j \alpha} + H.c.) 
    - \mu \sum \limits_{j} \sum \limits_{\alpha=1}^{4} n_{j \alpha} \nonumber\\
    &+ \sum \limits_{j} \sum \limits_{\alpha = 1}^{4} \sum \limits_{\alpha' > \alpha}^{4} U_{\alpha \alpha'} n_{j \alpha} n_{j \alpha'} \nonumber\\
    &+ V_{ex} \sum \limits_{j} (c_{j 2}^{\dagger} c_{j 3}^{\dagger} c_{j 1} c_{j 4} + H.c.),
\end{align}
where $c_{i\alpha}^{\dagger}$ ($c_{i\alpha}$) is the fermionic creation (annihilation) operator for a particle with flavor $\alpha$ located on lattice site $i$ and $n_{i\alpha}=c_{i\alpha}^{\dagger}c_{i\alpha}$ is the corresponding number operator.
The hopping amplitude $t$ and chemical potential $\mu$ are equal for all flavors and lattice sites, while $U_{\alpha \alpha'}$ are elements of the symmetric density-density interaction matrix, defined by $U_{\alpha \alpha}=0$, $U_{12}=U_{34}=U$, $U_{13}=U_{24}=V_{dir}-V_{ex}$, and $U_{14}=U_{23}=V_{dir}$. The last term of the Hamiltonian describes the {\it spin-flip} process that can be associated with  {\it Hund's coupling} in solid-state materials. Below, we restrict our analysis to the case of repulsively interacting fermions, such that all nonzero matrix elements $U_{\alpha \alpha'}$ are positive (and, in particular, $V_{dir}\geq V_{ex}$).

We focus on thermodynamic properties of the model~(\ref{eq:HubbardHwVex}) at half filling ($n=2$), where the system is particle-hole symmetric.
The corresponding chemical potential is given by
\begin{equation}\label{eq.muhf}
  \mu_{\text{hf}} = (U+2V_{dir}-V_{ex})/2.
\end{equation}

Of particular interest is the SU(4)-symmetric system where all spin and orbital degrees of freedom play an identical role in~(\ref{eq:HubbardHwVex}),
meaning that all interspecies interactions are equal to $U$ (i.e., $V_{dir}=U$ and $V_{ex}=0$). 
We start our discussion with this special case in Sec.~\ref{Results-SU4}.
The SU(4) symmetry is lowered as soon as any interactions become unequal.
We study two particular cases of the four-component Hubbard model with lower symmetries in Sec.~\ref{Results-brokenSU4}: First, we explore the role of interorbital direct interactions without taking the exchange interaction into account, i.e., $0<V_{dir}<U$, $V_{ex}=0$. 
Second, we consider a finite exchange interaction and include the spin-flip term.

We use DMFT, a numerical approach based on a mapping of the original lattice problem onto an effective Anderson impurity model~\cite{Georges1996RMP}. To solve the impurity problem, we mostly employ the exact-diagonalization (ED) solver~\cite{Caffarel1994PRL} since it is fast and reliable in most regimes of interest.
Moreover, building upon the generalized version for multicomponent mixtures~\cite{Sotnikov2015PRA}, it can be extended to account for the spin-flip term in a straightforward way (see the Appendix~\ref{app.A} for more details).
In places, a continuous-time quantum Monte Carlo hybridization expansion solver (CT-HYB) in the segment representation~\cite{Gul2011RMP,Sotnikov2014PRA} is used to benchmark the accuracy of the obtained results for the SU(4)-symmetric system.

We consider a simple cubic lattice that is directly related to experimental realizations with ultracold atoms in optical lattices.
The hopping amplitude~$t$ is used as a unit of energy throughout the paper (the bandwidth for the noninteracting system is $W=12t$).
Our description suggests that the system is homogeneous (but breaking of the lattice translation symmetry into two sublattices is possible) and infinite, such that the results apply to bulk properties of trapped gases or other materials only.

\section{Results}

\subsection{SU(4)-symmetric system} \label{Results-SU4}

\begin{figure}
\includegraphics{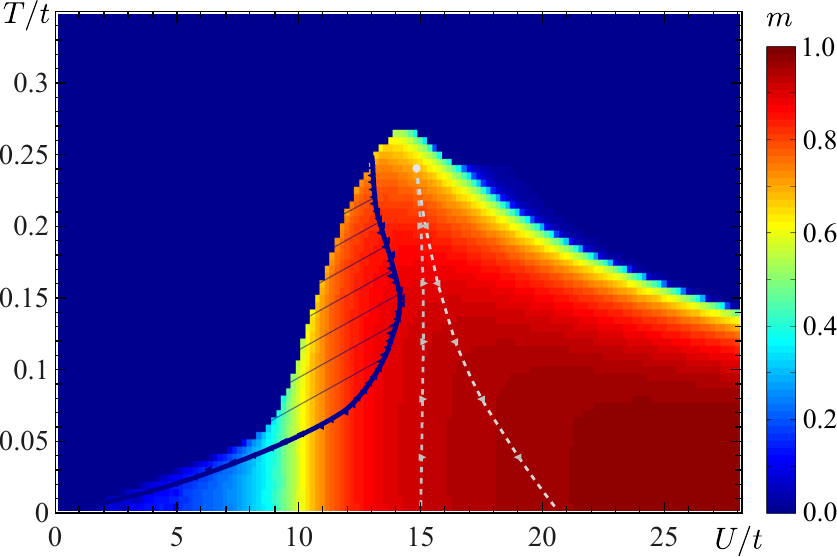}
\caption{\label{fig:pd} Phase diagram of the SU(4)-symmetric Hubbard model at half filling ($n=2$). The hatched area corresponds to the coexistence region of PM Fermi liquid and AFM insulator. The metal-insulator coexistence region [obtained under the PM constraint, i.e., no broken SU(4) symmetry] is indicated with light dotted lines inside the AFM phase.}
\end{figure}

We first focus on ordered phases in the SU(4)-symmetric system. 
The term ``magnetic'' is used to refer to the two-orbital Hubbard model.
At half filling (i.e., two particles per site), we expect that antiferromagnetic (AFM) correlations can develop under appropriate conditions.
In order to identify AFM ordered phases and to analyze their stability at finite temperature, we calculate the staggered order parameters $m_{\alpha}$ defined as
\begin{equation}
m_{\alpha} = |\tilde{m}_{\alpha}| \equiv |n^{A}_{\alpha} - n^{B}_{\alpha}|,
\end{equation}
where $n^{\gamma}_{\alpha}$ denotes the filling of flavor $\alpha$ on a site of sublattice $\gamma \in \{A, B\}$. Since we are dealing with fermions, the value of $m_{\alpha}$ ranges between 0 (paramagnet, PM) and 1 (``perfect'' AFM).
We observe that in the symmetry-broken phases, the four flavors always split up into pairs (see also reasoning below), with each pair dominantly occupying one of two sublattices. Staggered magnetizations are equal for members of each pair $\alpha \alpha'$ ($\tilde{m}_{\alpha} = \tilde{m}_{\alpha'}$) and opposite  for members of different pairs ($\tilde{m}_{\alpha} = -\tilde{m}_{\beta}$), but all of them have the same amplitudes, $m_{\alpha} \equiv m~\forall~\alpha$. Therefore, the AFM phase can be described by a single parameter $m$.
According to Fig.~\ref{fig:pd}, we find this phase to remain most stable against 
thermal fluctuations at intermediate interaction strength ($U\sim W=12t$).
The N\'{e}el temperature reaches its maximum $T_{\text{N}}^{\text{max}} \approx 0.27\,t$ at $U \approx 14.2\,t$ and decreases at larger interaction strengths according to the relation $T_{\text{N}} \propto t^2/U$ for the strong-coupling regime. 

Note that in our numerical analysis we applied no additional constraints on the type of magnetic order, except for limiting ourselves to the easy-axis (number-operator basis) projections and allowing only two distinct sublattice solutions.
However, in two particular regions of the $T-U$ phase diagram (see Fig.~\ref{fig:pd}), (i) $U\approx5t$, $T<0.13t$ and (ii) $U\approx14t$, $T\approx0.3t$, a damping between DMFT iterations was required to ensure final convergence, independently of the impurity solver (ED or CT-HYB). With an additional linear-response analysis, we verified that in case~(i), there are no other incommensurate competing types of magnetic instabilities [i.e., other than with the ordering wave vector ${\bf Q}=(\pi,\pi,\pi)$]. The absence of DMFT convergence without damping in region~(ii) is caused by the proximity of the metal-insulator transition.

\begin{figure}
\includegraphics{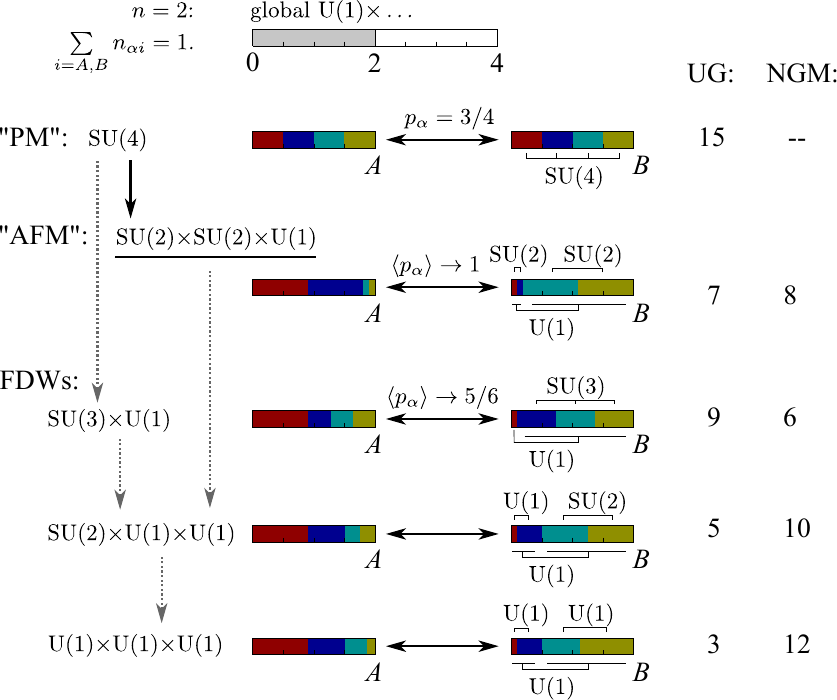}
\caption{\label{fig:phases} Sketch of some potential symmetry-broken phases on bipartite lattices with corresponding residual symmetries, the total number of unbroken generators (UG), and the number of Nambu-Goldstone modes (NGM) that is equal to the number of spontaneously broken generators at half filling~\cite{Caz2014RPP}. For better visibility, we used the number-operator-basis representation, so that occupancies of sublattices $A$ and $B$ by different components are indicated by different colors. $p_\alpha$ represents the probability of tunneling of component $\alpha$ between sublattices, which is determined according to the Pauli blocking principle.}
\end{figure}

The used approach allows, in principle, for observation and analysis of other types of long-range ordered states, in particular, other types of \emph{flavor-density waves} (FDWs) that have different residual symmetries and thus different number of unbroken generators and Nambu-Goldstone modes; see also Fig.~\ref{fig:phases}. 
However, while susceptibilities that correspond to other generators of the SU(4) group show equal behavior in the PM region with corresponding divergence at the same critical temperature, we observe that the system chooses the ``conventional'' AFM symmetry-broken phase at \emph{any nonzero coupling} $U/t$ when the temperature is decreased. This can be explained by the fact that the residual symmetry of the denoted AFM phase (in contrast to other FDWs) is maximally adjusted to Pauli blocking under given constraints, such that the kinetic energy of all four flavors can be minimized with the most efficiency.

\begin{figure}
\includegraphics{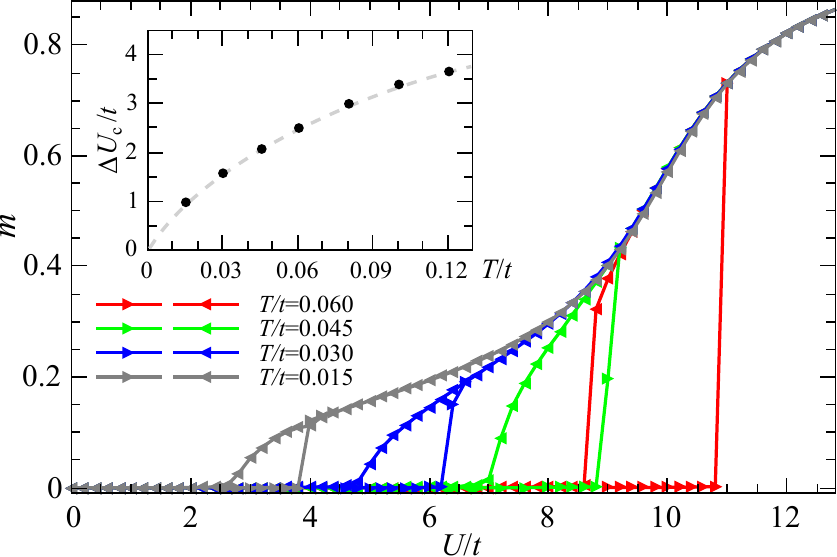}
\caption{\label{fig:add} Dependencies of the magnetization~$m$ on the interaction strength~$U/t$ at different temperatures. Inset: The width of the coexistence region as a function of the temperature.}
\end{figure}

The phase diagram presented in Fig.~\ref{fig:pd} for the SU(4)-symmetric system is peculiar in several aspects.
From the Fermi-liquid (weak-coupling) side, we observe a discontinuity in $m$, indicating a first-order phase transition from PM to AFM.
This result is in stark contrast to the well-studied case of the two-component SU(2)-symmetric Hubbard model at half filling, where the corresponding transition is of the second order at any coupling strength $U/t$. 
It also differs substantially from the low-temperature characteristics of the three-component SU(3)-symmetric Hubbard model on a simple cubic lattice at $n=1$ (1/3 band filling), where the transition is of the first order, but appears only at a moderate coupling $U_{c}\approx9.6t$ in the $T=0$ limit~\cite{Sotnikov2015PRA}.
In Fig.~\ref{fig:add}, we analyze in more detail the low-temperature behavior of the staggered magnetization~$m$ in the weak-coupling region of the obtained phase diagram (see Fig.~\ref{fig:pd}). Taking into account the extrapolation shown in the inset, we conclude that the coexistence region shrinks monotonously with temperature; thus the transition becomes second order only at $U=0$ and $T=0$.

\begin{figure}
\includegraphics{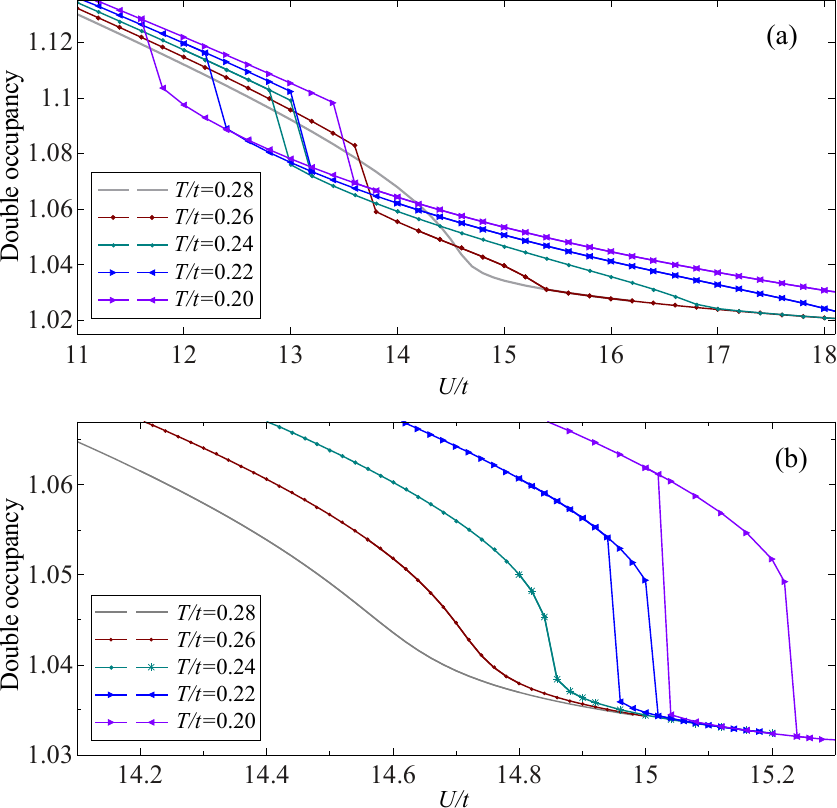}
\caption{\label{fig:PMhyst} Dependencies of the double occupancy on temperature (a) with account for AFM ordering and (b) under the PM constraint. Plots (a) and (b) also correspond to two different coexistence regions indicated in Fig.~\ref{fig:pd}.}
\end{figure}

The observed first-order transition is also accompanied by a hysteresis of the average double occupancy (see Fig.~\ref{fig:PMhyst}), $D = L^{-1} \sum_j D_j$, where $D_j = \sum_{\alpha} \sum_{\alpha' > \alpha} \langle n_{j \alpha} n_{j \alpha'} \rangle$ for a lattice site $j$ and $L$ is the total number of sites in the system. 
Therefore, both $m$ and $D$ can be used to determine the coexistence region of the PM Fermi liquid and AFM insulator (the hatched area in Fig.~\ref{fig:pd}). The structure of the phase diagram agrees well with the recent DMFT results obtained for a similar SU(4)-symmetric system with the Bethe lattice geometry~\cite{Yanatori2016PRB}.
Furthermore, based on the hysteretic behavior of double occupancy in the PM regime (with AFM ordering being artificially suppressed in the numerical procedure) we find a first-order metal-insulator transition (MIT)~\cite{Inaba2005PRB,Blumer2013PRB,Yanatori2016PRB}, indicated in the phase diagram in Fig.~\ref{fig:pd} by light dotted lines.

\begin{figure}
\includegraphics{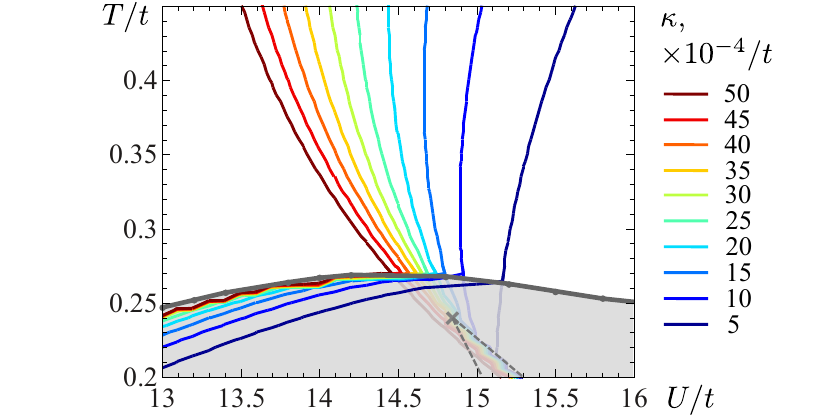}
\caption{\label{fig:compr} Contour plot for compressibility of the SU(4)-symmetric mixture at half filling in the proximity of the magnetic (thick gray line) and metal-insulator (dashed gray line) transitions. The shaded contour lines in the AFM region correspond to solutions under the PM constraint in DMFT.}
\end{figure}

The isothermal compressibility $\kappa$, defined as the variation of the particle density with the chemical potential, $\kappa n^2 = \partial n/\partial \mu$, gives further information about the single-particle gap. The advantage of this quantity is that it is both experimentally measurable~\cite{Schneider2008S} and theoretically obtainable, similarly to the double occupancy. Diverging lines of constant compressibility in Fig.~\ref{fig:compr} mark the metal-insulator crossover region above the AFM phase. 
Given that DMFT overestimates the N\'{e}el temperature in the intermediate- and strong-coupling regimes for the single-band SU(2)-symmetric Hubbard model (see, e.g., Ref.~\onlinecite{Kent2005PRB} for comparison), we expect that the MIT second-order critical point in the SU(4)-symmetric system lies in the PM region and can thus be directly probed in experiment. In the ordered phase, the MIT is shifted towards smaller interaction strengths and coincides with the AFM transition line. This effect is intuitively clear since at weak and intermediate coupling the AFM order drives the system to the insulating state, and thus suppresses charge (particle-number) excitations.

\begin{figure*}
\includegraphics[width=0.8\linewidth]{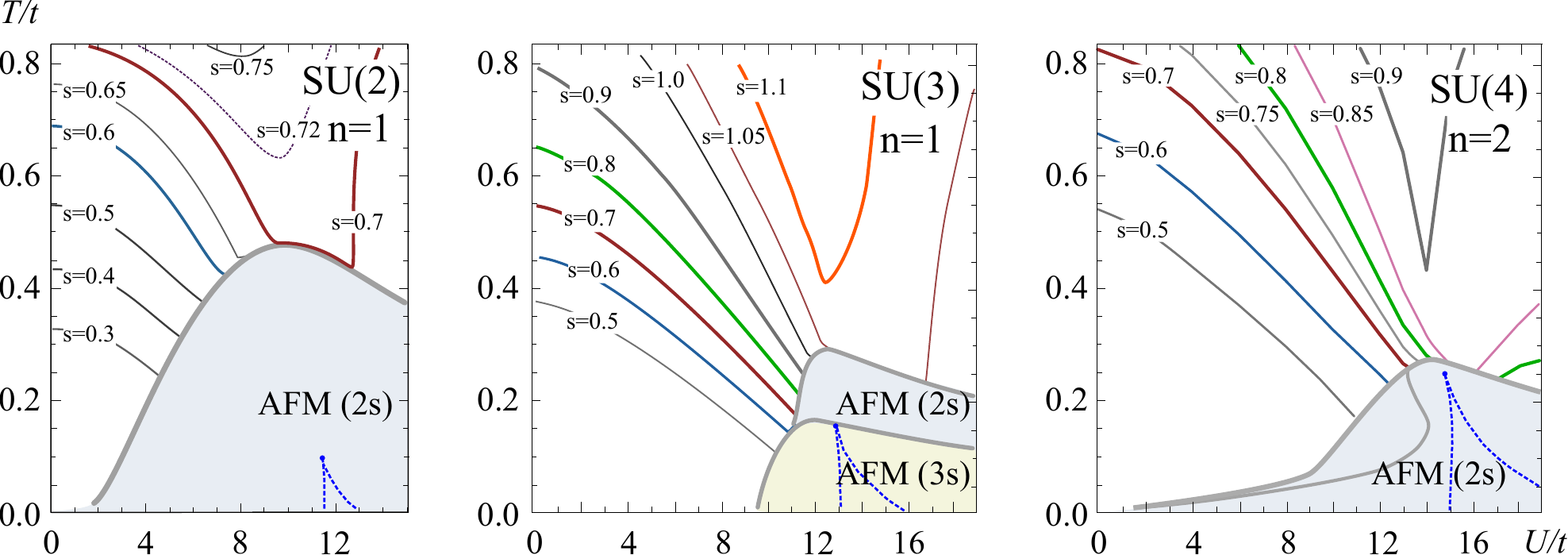}
\caption{\label{entrSU234} Isentropic lines indicating entropy per particle, AFM phases, and PM metal-insulator coexistence regions (blue dashed lines) for three different SU($N$)-symmetric Hubbard models on a cubic lattice. For comparison purposes, the subplots for two-component SU(2)-symmetric and three-component SU(3)-symmetric systems are taken from Refs.~\onlinecite{Sotnikov2012PRL} and \onlinecite{Sotnikov2015PRA}, respectively. Note that for the four-component mixture, the lattice filling is fixed to $n=2$, and thus the entropies per site are twice as high in this regime.}
\end{figure*}

The entropy per particle serves as a temperature measure in experiments with ultracold gases~\cite{Esslinger2010ARCMP, Haz2012PRA}.
We calculate the entropy per site in the normal (PM) phase and obtain $S(\mu_0, T)$ at a given temperature $T$ and chemical potential $\mu_0=\mu_{\text{hf}}$ (\ref{eq.muhf}) via the thermodynamic Maxwell relation $\partial S/\partial \mu =\partial n/\partial T$ by integration $S(\mu_0, T) = \int_{- \infty}^{\mu_0} \left(\partial n/\partial T \right) d\mu$. The entropy per particle for a homogeneous system is then determined as $s=S/n(\mu_0)=S/2$. According to Fig.~\ref{entrSU234}, in the region of weak and intermediate coupling, the well-known Pomeranchuk cooling effect is observed, i.e., at fixed entropy the temperature decreases with increasing $U/t$. For the SU(4)-symmetric mixture at half filling, the critical entropy value at which the isentropic curve reaches the AFM phase boundary is estimated to be $s_c \approx 0.86$.

From Fig.~\ref{entrSU234}, it appears most favorable to approach the N\'{e}el-type magnetic ordering with three-component mixtures in the case of homogeneous systems. However, under experimental conditions, the presence of the trapping potential plays a crucial role. 
%
There are strong indications that the exceeding entropy can be more effectively distributed in surrounding shells of four-component mixtures than in systems consisting of only two or three interacting fermionic flavors.  
The corresponding DMFT analysis taking account of the trapping potential can be done (see, e.g., Ref.~\onlinecite{Sotnikov2016PLA}); however, quantitative results depend on parameters of a particular experimental setup and thus this task goes beyond the scope of the present paper.
In Sec.~\ref{Results-brokenSU4}, we provide another comparison of entropic behavior in the context of four-component mixtures with lower symmetry of the Hamiltonian~(\ref{eq:HubbardHwVex}).

\subsection{Four-component mixtures with lower symmetries} \label{Results-brokenSU4}
\begin{figure}
\includegraphics{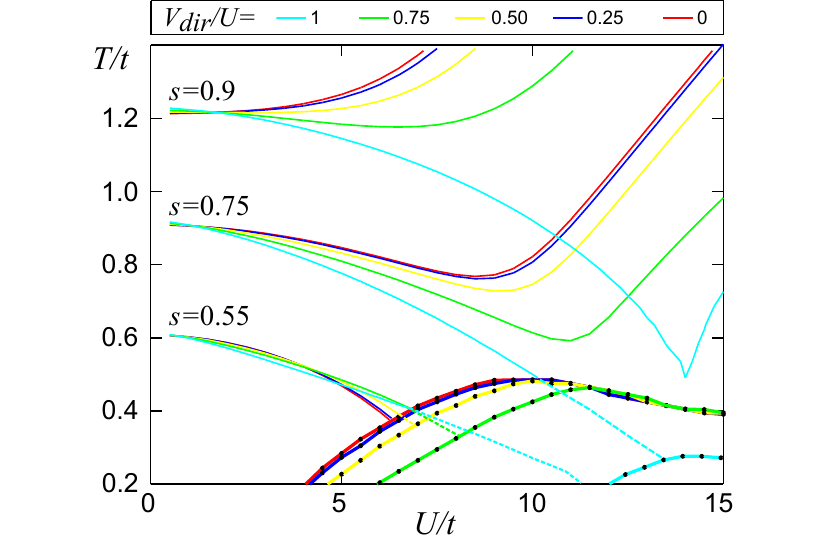}
\caption{\label{fig:VdirS} Isentropic lines and AFM phase boundary at different values of $V_{dir}/U$.}
\end{figure}

Next, we study the influence of the interorbital interaction~$V_{dir}$ by gradually decreasing its strength from $V_{dir}=U$ to $V_{dir}=0$ (while setting $V_{ex}=0$ in the first part of this section) and analyzing the transition from a complete SU(4) symmetry to the case of two fully separated (mutually noninteracting) SU(2)-symmetric systems (see Fig.~\ref{fig:VdirS}).
We observe that the interorbital interaction suppresses the AFM ordered phase. 
With a decrease of $V_{dir}$, this phase is enlarged significantly [the most rapid change is observed at $V_{dir}\approx U$, i.e., close to the SU(4)-symmetric point], both into the region of higher temperatures and lower interaction strengths. 
Note that the present AFM phase in the $V_{dir} = 0$ limit is identical to DMFT results obtained earlier for the single-band SU(2)-symmetric system in simple cubic lattice geometry~\cite{Fuchs2011PRL,Sotnikov2012PRL,Golubeva2015PRA}.

In contrast to temperature dependencies, the entropic behavior in the PM region demonstrates that the entropy capacity increases with the interorbital direct interaction strength. In particular, for $s = 0.9$, the lowest temperature that can be reached by adiabatic change of the interaction strengths is approximately $T=0.5t$ for $V_{dir} = U$, which is twice as large as for other cases depicted in Fig.~\ref{fig:VdirS}. Therefore, the critical entropy that allows one to approach the AFM ordered phase within the adiabatic process is significantly larger in the case of the SU(4)-symmetric fermionic mixture. Note that the SU(4) symmetry does not necessarily have to be exact in order to produce the increased entropy capacity. Therefore, ultracold mixtures of alkali atoms ($^6$Li and $^{40}$K as alternatives to $^{87}$Sr and $^{173}$Yb) can be appropriate candidates for approaching the magnetically ordered phases. 

Finally, we study the influence of the exchange interaction and consider the case $V_{dir}=U/2$ and $V_{ex}\geq0$. Note that the parameter $V_{ex}$ sets the amplitude of the spin-flip process, but it also appears in the density-density term, since it enters the interaction matrix $U_{\alpha \alpha'}$. Below, we consider these contributions separately, i.e., we analyze the effect of $V_{ex} > 0$ as a density-density interaction without including the spin-flip term (also denoted as {\it Ising-type Hund's coupling}, the form that was used, e.g., in Refs.~\onlinecite{Held1998EPJ,Cichy2016PRA} to analyze ferromagnetic instabilities away from half filling), as well as the system with a full account of the spin-flip term. We present results for two particular nonzero values of the exchange interaction: $V_{ex}=U/4$ (fulfilling the relation $V_{dir} = U - 2 V_{ex}$, typically applied in solid-state theory) and $V_{ex}=U/2$ (due to $V_{dir}=U/2$, this limit corresponds to zero off-diagonal elements $U_{13}=U_{24}=0$ in the density-density interaction matrix $U_{\alpha\alpha'}$).

\begin{figure}
\includegraphics{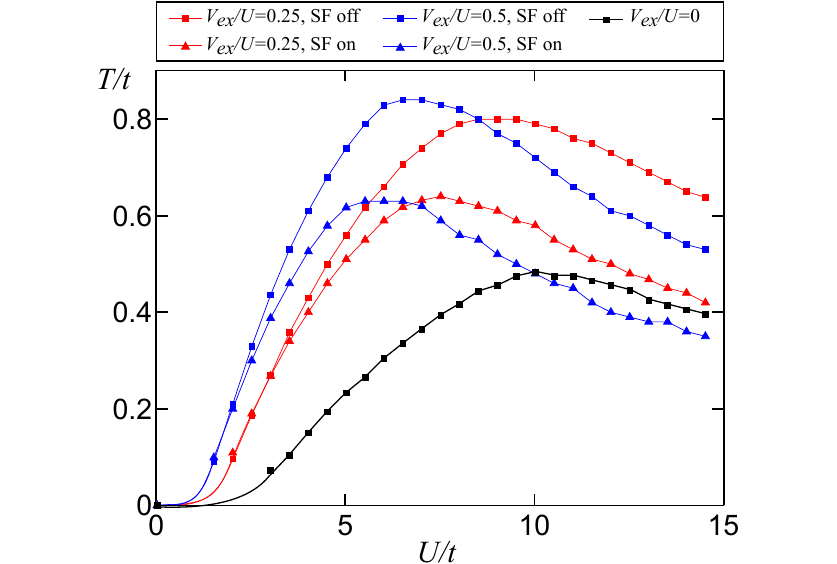}
\caption{\label{fig:sf} Phase transition lines to the AFM state for systems with $V_{dir}=U/2$ and $V_{ex}=0$ (black), $V_{ex}=U/4$ (red), or $V_{ex}=U/2$ (blue), with  the spin-flip term ($\blacktriangle$) and without it (i.e., only with the Ising-type contribution to the exchange, {\tiny $\blacksquare$}).}
\end{figure}
As shown in Fig.~\ref{fig:sf}, the AFM ordering remains a dominant instability for the Hubbard model for the exchange interaction $0\leq V_{ex}\leq U/2$ at half filling. Moreover, because of different amplitudes for intra- and interspin interactions, it results in the relative enlargement of the AFM phase in the corresponding $T-U$ diagram in both cases (with and without the spin-flip term).
As expected, the spin-flip term does suppress the AFM ordered phase; however, the AFM-favoring effect of the finite exchange interaction outweighs the suppression at weak and intermediate coupling, such that the maximal critical temperatures at $V_{ex}\gtrsim U/4$ are still about 30\% higher than in the system with $V_{ex}=0$. 

We have entropy-based estimates concerning the ``optimal'' type of the exchange term (with or without spin flip as considered above) that can be realized 
in experiments working with mixtures of ultracold atoms. Similarly to the above-discussed cases, we observe that 
at fixed $U$ and $T$ the entropy per particle increases (the corresponding Pomeranchuk effect becomes stronger) with the account of the spin-flip process in the proximity of magnetic transitions. 
Therefore, at intermediate coupling (close to the corresponding $T_{\text{N}}$ maxima), the two effects --- suppression of $T_{\text{N}}$ and increase in $s$ --- almost compensate each other, such that the critical entropies per particle are approximately the same, e.g., for $V_{ex}=U/4$, we obtain $s_c\approx0.67$ in both cases.

\section{Conclusions and Outlook}

We analyzed the thermodynamic properties of four-component fermionic mixtures in a periodic lattice with simple cubic (three-dimensional) geometry at half filling.
The DMFT results for SU(4)-symmetric mixtures (similarly to those obtained for the Bethe lattice~\cite{Yanatori2016PRB}) show a first-order transition from a paramagnetic Fermi liquid to an antiferromagnetic insulator at nonzero temperature, in contrast to the continuous transition in the two-component Hubbard model with SU(2) symmetry.
Another feature distinguishing the SU(4)- from SU(2)-symmetric model is the proximity of the Mott critical point to the AFM boundary.
%
%
The breaking of the SU(4) symmetry by $V_{dir}$ and $V_{ex}$ is found to increase the AFM critical temperature, but to reduce the critical entropy, which is the actual control parameter in ultracold-atom experiments.

The inhomogeneity and finite-size effects originating from the trapping potential are important and should be studied in more detail. 
Other important directions for a separate theoretical analysis are magnetic instabilities away from half filling as well as a special case of quarter filling in the four-component Hubbard model at nonzero temperature.

\appendix
\section{Modifications of the applied ED impurity solver}\label{app.A}

In order to understand which modifications to the ED solver are necessary, we analyze the effect of the spin-flip term on the basis states. The Anderson impurity model (AIM) corresponding to Eq.~(\ref{eq:HubbardHwVex}) is
\begin{eqnarray}\label{Haim}
 {\cal H}_{\text{AIM}} = \sum \limits_{l = 1}^{n_s} \sum \limits_{\alpha = 1}^{4} \varepsilon_{l \alpha} n_{l\alpha}
    + \sum \limits_{l = 2}^{n_s} \sum \limits_{\alpha = 1}^{4} V_{l \alpha} (a_{l \alpha}^{\dagger} c_{\alpha} + H.c.)\nonumber
    \\
    + \sum \limits_{\alpha = 1}^{4} \sum \limits_{\alpha' > \alpha}^{4} U_{\alpha \alpha'} n_{\alpha} n_{\alpha'} 
    + V_{ex}~(c_{2}^{\dagger} c_{3}^{\dagger} c_{1} c_{4} + H.c.), ~~ 
\end{eqnarray}
where all lattice sites except for the impurity ($l=1$, $\varepsilon_{1 \alpha}=-\mu$, $c_{\alpha}\equiv a_{1\alpha}$) constitute the effective bath with orbital structure. 
The number of orbitals taken into account, $n_s$, determines the accuracy of the model. In the Fock representation, the basis states read:
\begin{equation*}
\ket{ n_{1 1} n_{2 1} ... n_{n_s 1} } \ket{ n_{1 2} ... n_{n_s 2} } \ket{ n_{1 3} ... n_{n_s 3} } \ket{ n_{1 4} ... n_{n_s 4} }
\end{equation*}
with $n_{l \alpha} \in \{0, 1\}$ being the occupation number for flavor $\alpha$ on orbital $l$, where $l=1$ is the impurity and $l \geq 2$ denote the bath orbitals.
The basis states of the system can be grouped in sets labeled by the configuration ${\bf q} = (q_{1}~ q_{2}~ q_{3} ~q_{4})$, with $q_{\alpha} = \sum_{l=1}^{n_s} n_{l \alpha}$ denoting the total number of particles with flavor $\alpha$ in the system. 

When the spin-flip term is absent, $q_\alpha$ is a conserved quantity and these sets build separate blocks in the block-diagonal Hamiltonian.
The total number of particles $n = \sum_{\alpha} q_{\alpha}$ in the system can vary between 0 ($q_{\alpha} = 0~\forall~\alpha$) and $N \cdot n_s$ ($q_{\alpha} = n_s~\forall~\alpha$), where $N$ is the number of flavors. 
In total, there are $X$ distinct configurations $\bf{q}$,
\[
  X=(n_s+1)^N,
\]
and to each configuration $\bf{q}$ belong $Y$ basis states building one block in the Hamiltonian, with
\[
Y = \prod \limits_{\alpha} \frac{ n_s! }{ (n_s-q_{\alpha})!~q_{\alpha}!} \equiv \prod \limits_{\alpha} \binom{n_s}{q_{\alpha} }.
\]

The spin-flip term does not change the total number of particles in the system; it only alters the configuration $\bf{q}$ and thus connects different $\bf{q}$ blocks in the Hamiltonian. A block associated with configuration $\bf{q}$ becomes connected to blocks of the following configurations:
\[
(q_{1}~q_{2}~q_{3}~q_{4}) \rightarrow \begin{cases} (q_{1} + b \quad q_{2} - b \quad q_{3} - b \quad q_{4} + b) \\
								    ... \\
								    (q_{1} + 1 \quad q_{2} - 1 \quad q_{3} - 1 \quad q_{4} + 1) \\
								    (q_{1} \hspace{1.15cm} q_{2} \hspace{1.15cm} q_{3} \hspace{1.15cm} q_{4}) \\
								    (q_{1} - 1 \quad q_{2} + 1 \quad q_{3} + 1 \quad q_{4} - 1) \\
								    ... \\
								    (q_{1} - a \quad q_{2} + a \quad q_{3} + a \quad q_{4} - a) \end{cases}
\]
where $a = \min{[ q_{1}, n_s-q_{2}, n_s-q_{3}, q_{4} ]}$ and $b = \min{[ n_s-q_{1}, q_{2}, q_{3}, n_s-q_{4} ]}$. In this way, the dimension of the block $i$ to diagonalize increases from $Y_i \times Y_i$ with
$
Y_i = \prod_{\alpha} \binom{n_s}{q_{\alpha i}}
$
to $B_j \times B_j$ with
$
B_j = \sum_{i} Y_i
$
for the $j$-th bunch of $n$ blocks $i_1,\ldots,i_n$, which are connected inside the bunch $j$ by the spin-flip term.


Therefore, in order to account for the spin-flip term in the ED solver while preserving the efficiency, it is sufficient to regroup the basis states, such that all connected configurations are adjacent. Thus, the Hamiltonian preserves a block-diagonal structure, only with a fewer number but larger size of the blocks.

Regarding limitations in the accuracy of the ED solver originating from the finite number of orbitals $n_s$ in Eq.~(\ref{Haim}),
from a direct comparison with the CT-HYB solver we determined that $n_s=3$ ($n_s=4$) orbitals per each of four fermionic flavors is usually enough to have a reliable qualitative (quantitative) agreement in most regimes of interest. Therefore, in our analysis we use mainly $n_s=4$ (in particular, in Sec.~\ref{Results-SU4}), except in cases that require qualitative estimates, but significantly enlarge the parameter space to be analyzed (results with a full account of the spin-flip process and entropy calculations that are given in Sec.~\ref{Results-brokenSU4}).

\begin{acknowledgments}
The authors thank A. Koga and P. van Dongen  for fruitful discussions.
A.G. acknowledges support by the Perimeter Institute for Theoretical Physics. Research at the Perimeter Institute is supported by the Government of Canada through the Department of Innovation, Science and Economic Development Canada and by the Province of Ontario through the Ministry of Research, Innovation and Science.
A.S. and J.K. acknowledge funding of this work from the European Research Council (ERC) under the European Union's Horizon 2020 research and innovation programme (Grant Agreement No. 646807-EXMAG).
W.H. acknowledges financial support from the Deutsche Forschungsgemeinschaft via DFG SFB/TR 49, DFG FOR 2414, DFG SPP 1929 GiRyd, and the high-performance computing center LOEWE-CSC.
\end{acknowledgments}

\bibliography{SU4}	
\end{document}